\documentclass{ws-procs9x6-cpt22}
\begin{document}

\newcommand{\refeq}[1]{(\ref{#1})}
\def\etal {{\it et al.}}

\title{Testing Discrete Symmetries in Ortho-Positronium Decays\\ with the J-PET Detector}

\author{E.\ Czerwi\'nski,$^{1,2}$}

\address{$^1$Marian Smoluchowski Institute of Physics, Faculty of Physics, Astronomy and Applied Computer Science, Jagiellonian University, Krak\'ow, 30-348, Poland}

\address{$^2$Center for Theranostics, Jagiellonian University, Krak\'ow, Poland}

\author{On behalf of the J-PET Collaboration}

\begin{abstract}
A first result related to a CPT test in the decay of ortho-positronium based on data collected by means of the J-PET detector is already available,\cite{Moskal:2021kxe} and 
a new test of CP symmetry is under investigation.\cite{Czerwinski:2022tvd}
Both results are based on the determination of angular correlations between momenta of gamma rays and
other vector properties of the investigated system.\cite{Adkins:2010pd}
Here, 
arguments for the possible usage of data collected by the J-PET group to test CPT symmetry within the Standard-Model Extension framework\cite{datatables}
are shortly presented.
\end{abstract}

\bodymatter

\section{Introduction}
The bound state of an electron and a positron (positronium) allows for tests of discrete symmetries 
either via the measurement of energy levels\cite{Kostelecky:2015nma} or via 
the determination of expectation values of symmetry-odd operators for specific symmetries.\cite{Adkins:2010pd}
For the second approach the previously investigated angular correlations for CPT\cite{Vetter:2003} and CP\cite{Yamazaki:2009hp}
symmetry involve momenta of gamma rays and the spin of ortho-positronium.
The unique properties of the J-PET detector allow for a new class of operators to be investigated.\cite{Moskal:2016moj}
Additionally the process of data collection is continuous for long periods and it's influence on the registered data is minimal.

\section{Experimental techniques}
The Jagiellonian Positron Annihilation Tomograph (J-PET) is optimized for the detection of annihilation gammas originating from electron--positron annihilation.\cite{Dulski:2020pqi}
Due to the usage of plastic scintillators as detection elements the timing properties are very good and the high granularity of the scintillating strips
allows for the identification of gamma scatterings.
The combination of these properties of the J-PET system directly translates into the unique features of the determination of the polarization directions of
the annihilation gammas on an event-by-event basis.\cite{Moskal:2016moj}

The relatively easy access to the central part of the detector provides an opportunity to use a different size of targets for positrons originating from
sodium sources. These targets increase the probability of ortho-positronium creation.\cite{Gorgol:2020acta}
A target of large size in the shape of a cylinder is used
to separate the point of positron emission (source position) and the annihilation point. A trilateration technique is used for the determination of the origin point
of the three gamma rays from ortho-positronium annihilation.\cite{Gajos:2016nfg} 
The velocity direction of the positron is inferred from its emission and annihilation points.
Since positrons created in $\beta$-decay are polarized, and this polarization is preserved to some extend after positronium creation, the measurement
of the velocity direction is equivalent to the determination of the positronium-spin direction. Again, this is applied on an event-by-event basis at J-PET.\cite{Moskal:2021kxe}

It is worth mentioning that studies of the following angular correlations have been performed at J-PET: a) between the spin direction and the annihilation plane, and b) between the polarization plane of a gamma and the momentum of one of remaining gamma rays. These studies show full angular coverage in terms of both, geometrical acceptance as well as analysis efficiencies.

\section{DAQ and data taking campaigns}
The data-acquisition system of J-PET operates in trigger-less mode.\cite{Korcyl:2016pmt,Korcyl:2018yjq}
Data are stored if a photomultiplier registers a signal above a small voltage threshold.
Therefore, multiple offline analyses dedicated to specific studies can be performed without being affected by trigger requirements.

Thus far the J-PET group conducted several continuous data-taking campaigns. The periods devoted to measurements for CPT-symmetry studies lasted
from 26 days\cite{Moskal:2021kxe} up to 9 months, being 16 months in total, while measurements for CP symmetry lasted from 6 days up to 8 months (12 months in total).

It is worth mentioning that new data-taking campaigns using a  modular version of the J-PET detector are scheduled already,
with the aim to improve the already published result.\cite{Neha:CPT22}

\section{Conclusions}
The properties of the J-PET detector, such as the determination of positronium spin direction and polarization plane of annihilation gammas on an event-by-event basis, together with trigger-less and continuous data taking for long periods are a good starting point to test CPT symmetry within the SME framework.  
Since the origin of time in the Sun-centered frame is set as the 2000 vernal equinox, the beginning of the J-PET local sidereal time is
$T_{\bigoplus|J-PET}=$11:40~PM~CET for all the data collected so far ($\lambda_{J-PET}=19.91\deg$).

\section*{Acknowledgments}
This work was supported by
the Foundation for Polish Science through the
TEAM POIR.04.04.00-00-4204/17 program,
the National Science Center of Poland through grants MAESTRO no.\
2021/42/A/ST2/00423 and OPUS no.\ 2019/35/B/ST2/03562,
the Ministry of Education and Science through grant no.\ SPUB/SP/490528/2021,
the EU Horizon 2020 research and innovation program,
STRONG-2020 project, under grant agreement no.\ 824093,
and the SciMat and qLIFE Priority Research Areas  budget under the program {\it Excellence Initiative - Research University} at Jagiellonian University,
and Jagiellonian University project no.\ CRP/0641.221.2020.

\end{document}